\documentclass[letterpaper]{article} % DO NOT CHANGE THIS
\usepackage{aaai23} % DO NOT CHANGE THIS
\usepackage{times} % DO NOT CHANGE THIS
\usepackage{helvet} % DO NOT CHANGE THIS
\usepackage{courier} % DO NOT CHANGE THIS
\usepackage[hyphens]{url} % DO NOT CHANGE THIS
\usepackage{graphicx} % DO NOT CHANGE THIS
\usepackage{natbib} % DO NOT CHANGE THIS AND DO NOT ADD ANY OPTIONS TO IT
\usepackage{caption} % DO NOT CHANGE THIS AND DO NOT ADD ANY OPTIONS TO IT
\usepackage{algorithm}
\usepackage{algorithmic}

\usepackage{newfloat}
\usepackage{listings}
\DeclareCaptionStyle{ruled}{labelfont=normalfont,labelsep=colon,strut=off} % DO NOT CHANGE THIS
\floatstyle{ruled}
\newfloat{listing}{tb}{lst}{}
\floatname{listing}{Listing}
\usepackage{amsmath}
\usepackage{amsfonts}
\usepackage[capitalize]{cleveref}
\usepackage{booktabs}
\usepackage{multirow}

\title{Decision-Making Context Interaction Network\\ for Click-Through Rate Prediction}
\author{
 Xiang Li\textsuperscript{\rm 1}\equalcontrib,
 Shuwei Chen\textsuperscript{\rm 1}\equalcontrib,
 Jian Dong\textsuperscript{\rm 1},
 Jin Zhang\textsuperscript{\rm 1},
 Yongkang Wang\textsuperscript{\rm 1},\\
 Xingxing Wang\textsuperscript{\rm 1},
 Dong Wang\textsuperscript{\rm 1}
}
\affiliations{
 \textsuperscript{\rm 1}Meituan, Beijing, China\\
 \{lixiang172,chenshuwei04,dongjian03,zhangjin11\}@meituan.com\\
 \{wangyongkang03,wangxingxing04,wangdong07\}@meituan.com
}

\usepackage{bibentry}
\begin{document}

\maketitle

\begin{abstract}
Click-through rate (CTR) prediction is crucial in recommendation and online advertising systems. Existing methods usually model user behaviors, while ignoring the informative context which influences the user to make a click decision, e.g., click pages and pre-ranking candidates that inform inferences about user interests, leading to suboptimal performance. In this paper, we propose a Decision-Making Context Interaction Network (DCIN), which deploys a carefully designed Context Interaction Unit (CIU) to learn decision-making contexts and thus benefits CTR prediction. In addition, the relationship between different decision-making context sources is explored by the proposed Adaptive Interest Aggregation Unit (AIAU) to improve CTR prediction further. In the experiments on public and industrial datasets, DCIN significantly outperforms the state-of-the-art methods. Notably, the model has obtained the improvement of CTR+2.9\%/CPM+2.1\%/GMV+1.5\% for online A/B testing and served the main traffic of Meituan Waimai advertising system. 
%Code will be released at https://github.com/DCIN-anonymous/DCIN.
\end{abstract}

\section{Introduction}
The performance of Click-through rate (CTR) prediction model has a direct impact on final revenue and user satisfaction, and is therefore critical for recommendation and advertising systems. 
In recent years, deep network has been introduced to CTR prediction due to its powerful modeling capability.
Feeding informative data to a deep network with carefully designed structure, it learns the most representative features for predicting and usually generalizes well.
\par Early CTR prediction models~\cite{wide_deep,deepfm,dcn,xdeepfm,autoint} have designed specific components to learn sophisticated low-/high-order interactions among different feature fields and obtained significant improvement, but they ignore the relation among the user interacted items.
Recently, some pioneering methods~\cite{din,dsin,sim,can} mitigate this problem by modeling user historical behaviors: the intrinsic properties contained in the items with positive user feedbacks (e.g., click) are modeled as user interests to enrich the information perceived by CTR prediction model.
Though significant improvement has been made, this paradigm still faces some crucial issues.
On the one hand, it only considers pointwise information about whether each behavior matches user's interests, while the historical behaviors tend to be noisy~\cite{kalman}, making the extracted user interests imprecise.
On the other hand, this paradigm only easily models the superficial information from user historical behaviors, while ignoring the latent yet valuable contexts in the system (e.g., items' co-occurrence relationship), leading to suboptimal performance.
We emphasize that there is a need to explore these contexts to learn more stable representations.

\begin{figure}
\centering
\begin{tabular}{cc}
\includegraphics[width=3.5cm,height=5cm]{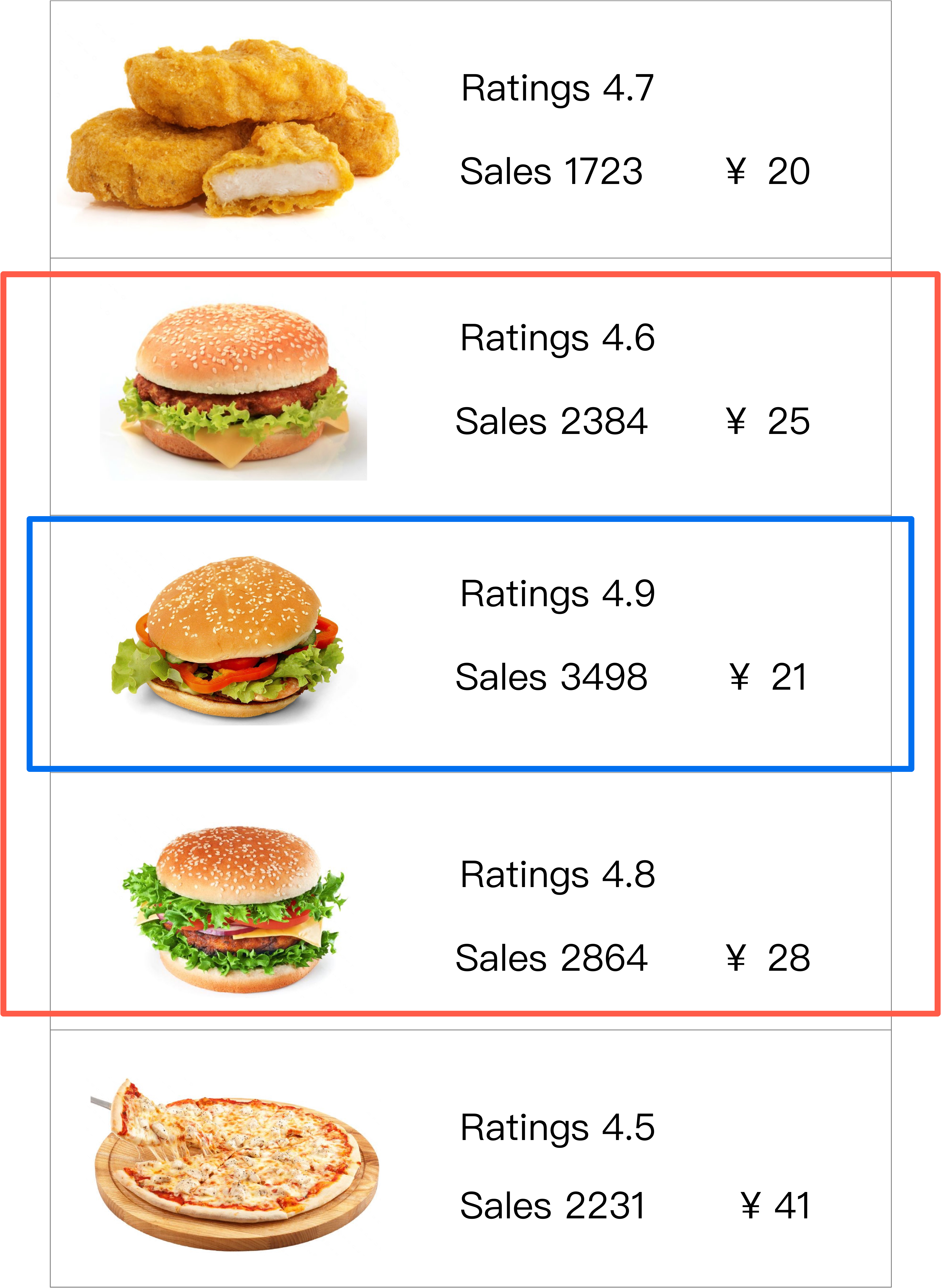} &
\includegraphics[width=3.5cm,height=5cm]{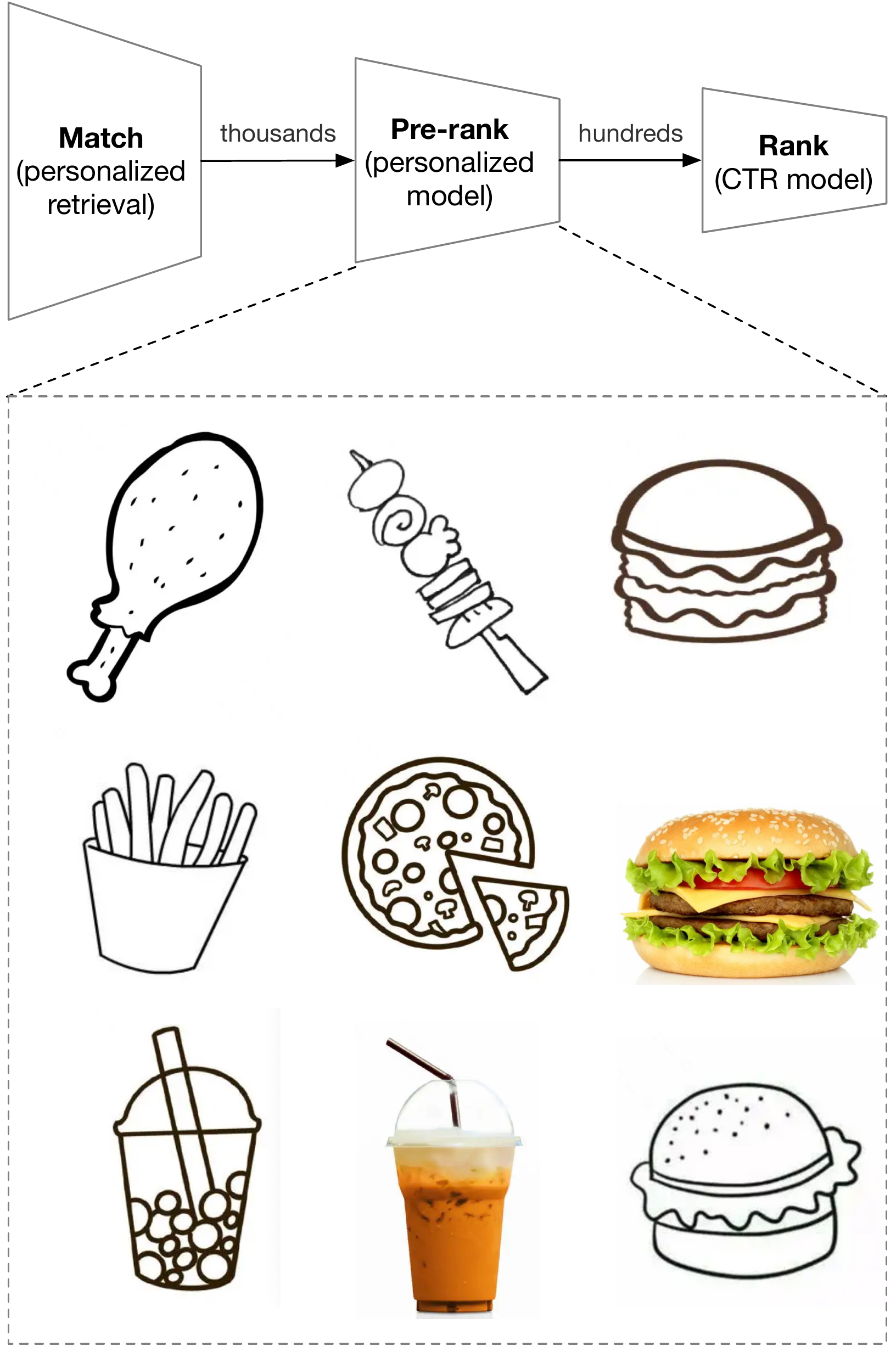} \\
 (a) explicit context & (b) implicit context
\end{tabular}
\caption{Illustration of explicit and implicit decision-making contexts. (a) The user proactively compares the attributes of the highly-related items in a local scope colored in red before making a click decision colored in blue. (b) We can deduce that the user likes fast food so much that the pre-ranking stage generates these candidates for him.}
\label{fig:decision_making_contexts}
\end{figure}

\par First, page-level co-occurrence relationship should be utilized.
In a typical display advertising system like Meituan Waimai (the largest food delivery platform in China) shown in~\cref{fig:decision_making_contexts}(a), a user swipes through the ad list to find his preferred items.
Suppose that the user likes eating fast foods and this personalized page with three burgers co-occurred was exposed to him earlier and he clicked the burger in the blue box, behavior modeling based methods use only the blue one to represente user interest without considering the intra-page items' mutual influence.
In fact, before deciding which item to click, the user concentrates on not only the blue one but also the other burgers in the local scope colored in red.
The user compares their attributes like prices, ratings, and sales explicitly, and then makes a click decision, so we name the relation of the intra-page items \textbf{explicit decision-making context}.
Users' comparison over some items showcases their interests on these items, so this explicit context and user behavior complement each other when inferring user interests.

\par Second, personalization in the system should be leveraged. Matching~\cite{covington2016deep,zhu2018learning} and pre-ranking~\cite{wang2020cold}, the two stages before ranking, take personalization into account to filter out items that may meet the user's interests.
The pre-ranking stage generates a set of candidates, that is, the target items to be predicted CTR in the ranking stage.
Due to personalization, there are many highly-related items in the set.
As illustrated in~\cref{fig:decision_making_contexts}(b), some fast foods are included in the candidates generated for the user in~\cref{fig:decision_making_contexts}(a).
When the CTR prediction target is the colored burger, we have more confidence that the user will click it because the presence of the other two burgers suggests that the user may like fast foods.
For the colored milk tea, since another milk tea exists, it may also have a higher CTR than a milk tea without any similar item in the candidates.
Though the users have not yet made click decision, the personalized results implicitly tell us what the users' interests might be and assist in prediction, so we name the information introduced by personalization \textbf{implicit decision-making context}.
\par Motivated by these observations, in this work, we propose a \textbf{D}ecision-Making \textbf{C}ontext \textbf{I}nteraction \textbf{N}etwork (DCIN), which simultaneously learns explicit and implicit decision-making contexts to make complete use of information in the system, enabling deep learning to unleash its capability:
\begin{itemize}
 \item \textbf{Explicit decision-making context modeling}.
 Considering that the user interest are not only latent in their click behaviors, but also in the explicit decision-making context, we thus augment user behavior interest with the explicit context.
 Specifically, for each item in user's click sequence, we first split out an exposure page that encloses it.
 Then the context in the page is utilized to augment click interest.
 However, not each page exactly contains this context especially those with extraneous items.
 To relieve this problem, we propose a \textbf{C}ontext \textbf{I}nteraction \textbf{U}nit (CIU), where the intra-page items are explicitly divided into two categories: items that are relevant/irrelevant to the clicked item. Then the behavior interest interacts only with explicit context from the relevant items, while the influence from the irrelevant ones is suppressed.
 \item \textbf{Implicit decision-making context modeling}.
 User interests drive the personalized system to generate some homogeneous candidates.
 Conversely, implicit decision-making context in these relevant candidates can be leveraged to deduce user interests.
 Therefore, we learn implicit contexts to refine target representation to highlight its attributes that may activate user interests.
 Meanwhile, the influence of irrelevant candidates should be suppressed.
 Thanks to the generality of the proposed CIU, we reuse it to interact target with the implicit context to produce a more comprehensive target representation.
\end{itemize}
\par Unfortunately, since the explicit and implicit contexts are modeled in isolation, there are two limitations in our formulation.
First, all the refined targets share the same augmented behavior interests.
Second, for a specific target, each behavior interest contributes equally.
The two limitations lead to inferior CTR prediction performance, as only partial interests should be activated when the user makes a click decision~\cite{din}.
To tackle this problem, we propose an \textbf{A}daptive \textbf{I}nterest \textbf{A}ggregation \textbf{U}nit (AIAU), in which the augmented behavior interests are adaptively aggregated according to their relevance w.r.t. the refined target.
\par In summary, our contributions are three-fold:
\begin{itemize}
 \item We propose a Decision-Making Context Interaction Network (DCIN), which learns explicit and implicit decision-making contexts simultaneously to unleash the capability of deep learning.
 To the best of our knowledge, we are the first to model both contexts in ranking stage.
 \item We propose a Context Interaction Unit (CIU) to effectively model the explicit and implicit contexts, and an Adaptive Interest Aggregation Unit (AIAU) is introduced to learn target-specific user behavior interests.
 \item Extensive experiments on Meituan Waimai's dataset validate our designs' effectiveness. Our model has been successfully deployed in the online display advertising system of Meituan Waimai, benefiting the improvement of the business.
\end{itemize}

\section{Related Work}
Click-through rate (CTR) prediction aims to predict the probability of a user clicking on the candidate item.
Early CTR predition methods mostly focus on capturing low-/high-order interactions of different feature fields.
Wide\&Deep~\cite{wide_deep} and DeepFM~\cite{deepfm} both deploy a wide component to model low-order interactions.
DCN~\cite{dcn} and xDeepFM~\cite{xdeepfm} model high-order interactions via explicit cross networks.
AutoInt~\cite{autoint} adopts self-attention~\cite{attn_need} to automatically learn high-order interactions. 
These pioneering works have demonstrated the capability of deep learning, but their performance saturates as they neglect the relation of user interacted items.
\par User behavior modeling based methods incorporate the highly personalized information across user interacted items into CTR prediction models.
DIN~\cite{din} introduces a local activation unit to extract user interest that is activated by the target.
DIEN~\cite{dien} adapts GRU~\cite{gru} to learn user interest evolution.
DSIN~\cite{dsin} leverages a bidirectional LSTM~\cite{lstm} to model intra-/inter-session user interest.
SIM~\cite{sim} introduces a cascaded search paradigm to model lifelong sequential behavior data.
CAN~\cite{can} proposes a Co-Action Network to fit complex feature interactions.
Though these methods show promising results, they fail to make full use of information in the system.
\par DFN~\cite{dfn} and DSTN~\cite{dstn} argue that user's negative feedbacks (e.g., unclick) also inform inferences about user interest.
However, the deployment of DSTN makes it more like a re-ranking model.
Some recent works have considered not only what information is fed to the model, but also what the structure of the information is, pushing forward the frontier of CTR prediction.
RACP~\cite{racp} and DPIN~\cite{dpin} model the entire exposure page around the user interacted item to learn stable user interest.
However, RACP models each page without regard to whether that page contains user feedback or valuable context, and DPIN is a re-ranking model.
CIM~\cite{cim} leverages the candidates generated by upstream relevance filter to represent user awareness, while ignoring the explicit behavior context.
Moreover, the impression probability of CIM averages a total number of 300 candidates, which may lead to feature smoothing and suboptimal result.
\par In this work, we not only focus on feeding adequate information into the deep model, but also propose modules that can filter out target-related information to ease learning and thus improve CTR predition performance.

\begin{figure*}[t]
\centering
\includegraphics[width=1.0\textwidth]{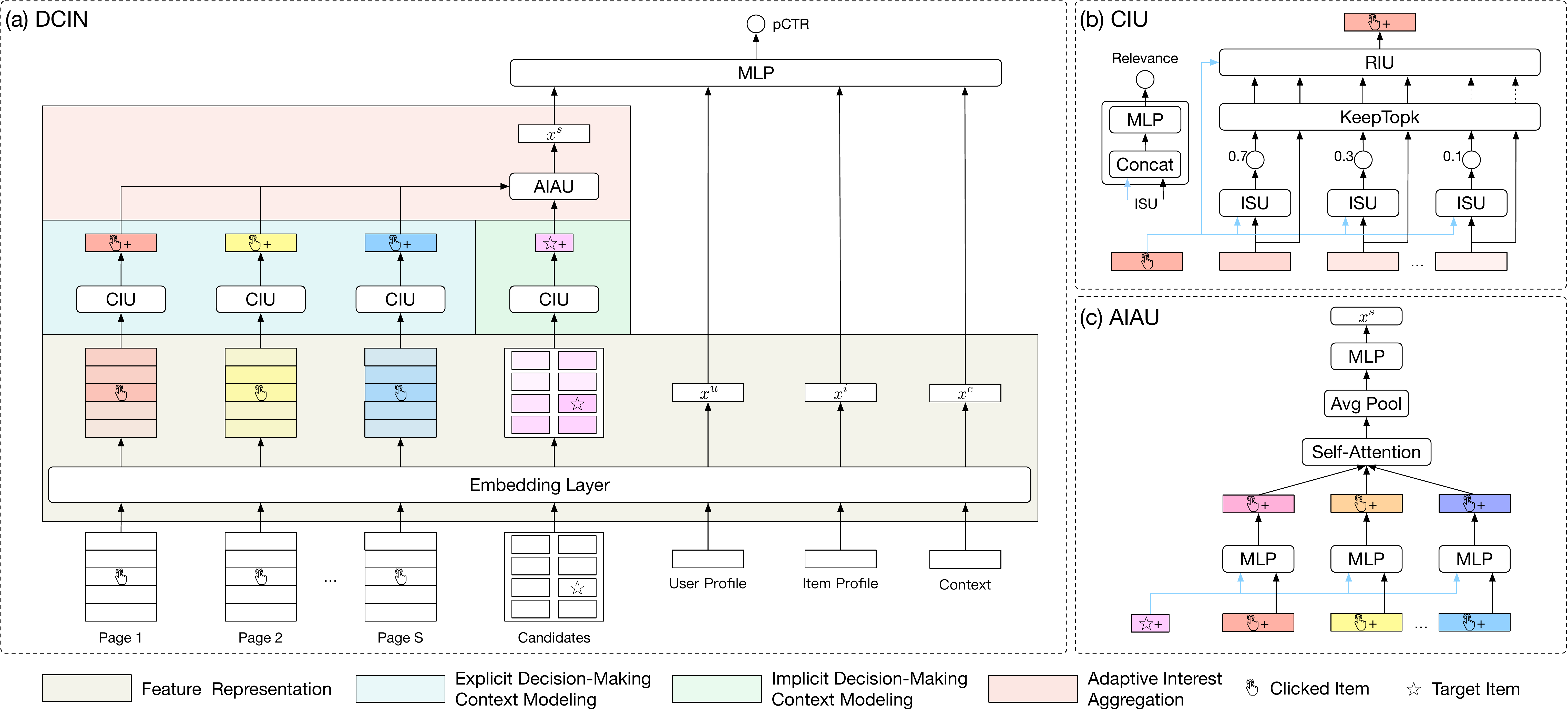}
\caption{Network architecture. (a) is our proposed DCIN, whose modules are painted in different colors. (b) shows the details of CIU, which is used to model explicit and implicit decision-making contexts. (c) shows the structure of AIAU, which is used to adaptively aggregate behavior interests that are relevant with the target item.}
\label{fig:dcin}
\end{figure*}

\section{Methodology}
Our aim is to predict the probability of a user clicking on the candidate items.
The overall architecture of DCIN is illustrated in~\cref{fig:dcin}(a).
It mainly contains five components: Feature Representation, Explicit Decision-Making Context Modeling, Implicit Decision-Making Context Modeling, Adaptive Interest Aggregation, and a final MLP for CTR prediction.

\subsection{Feature Representation}
The input features of CTR prediction model are mostly in a high-dimensional sparse form. They are usually transformed into low-dimensional dense representations via an embedding layer to reduce resource overhead and ease learning.
Our model takes user profile features, item profile features and context features as inputs, and transforms them into $x^{u} \in \mathbb{R}^{D_{u}}, x^i \in \mathbb{R}^{D_i}, x^c \in \mathbb{R}^{D_c}$, respectively. $D_u, D_i, D_c$ denote the embedding dimensions.
Besides, the user's click sequence and the corresponding pages, and the candidates generated by pre-ranking are input to the model:
\begin{itemize}
 \item \textbf{User's click sequence and pages}.
 User's click sequence and pages are used to model explicit decision-making context.
 The click sequence contains $S$ items clicked by the user.
 For the $i$-th clicked item, we split out a click page that encloses it with $P$ items.
 Each clicked item or page item contains features such as id, price, their position in the page, etc, and is transformed into a $D_s$-dimensional feature vector.
 The click sequence is represented as $xc = \{xc_1, xc_2, \dots, xc_{S}\}$, and the $i$-th click page is represented as $p_i = \{p_{i,1}, p_{i,2}, \dots, p_{i,P}\}$.
 \item \textbf{Candidates}. 
 We introduce candidates generated by pre-ranking stage to model implicit decision-making context.
 Each of the $C$ candidates contains features such as id, price, etc.
 The representation of the candidates is $xs=\{xs_1, xs_2, \dots, xs_C\}$, and $xs_i \in \mathbb{R}^{D_t}$.
 Note that the target item $xt \in \mathbb{R}^{D_t}$ is one of the candidates.
\end{itemize}

\subsection{Explicit Decision-Making Context Modeling}
User's explicit comparison on the intra-page items before making click decision provides crucial clues to infer user interests.
However, how to utilize this context remains challenging, because the user does not compare on every page. For example, on the pages with extraneous items, user basically clicks based on pointwise interests without comparison. 
To tackle this challenge, we propose a Context Interaction Unit (CIU), in which the user behavior interest is augmented by adaptively interacting with the complementary explicit context.
\par As illustrated in~\cref{fig:dcin}(b), CIU contains two components: 
1) Irrelevance Suppression Unit (ISU), which is deployed to suppress the influence from the intra-page items that are irrelevant with the clicked item; 
2) Relevance Interaction Unit (RIU), in which user behavior interest is augmented by interacting with the explicit context of the relevant intra-page items.

\noindent \textbf{Irrelevance Suppression Unit}.
As the advertising system usually expose a variety of items to the user, some of them are not in the user's range of comparison, and thus do not contribute to explicit context modeling.
This unit aims to divide the intra-page items into two categories, i.e., items that are relevant/irrelevant with the clicked item, so as to preserve the most informative context from the relevant items while suppressing the influence from these irrelevant ones.
Given the $i$-th clicked item $xc_i$ and the corresponding click page $p_i = \{p_{i,1}, p_{i,2}, \dots, p_{i,P}\}$, we first formalize their relevance $f_i = \{f_{i,1}, f_{i,2},\dots, f_{i,P}\}$ as: 
\begin{equation}
f_{i,j} = \text{MLP} ([xc_i, p_{i,j}, xc_i-p_{i,j}, xc_i \odot p_{i,j}])
\end{equation}
where $[\cdot]$ denotes concatenation, $\odot$ denotes element-wise product, and the used MLP contains two hidden layers. Then the top-$k_1$ relevant items are preserved, while the others are considered irrelevant and are directly suppressed, facilitating the following context interacting process:
\begin{equation}
 f_{i,j} = \left\{
 \begin{aligned}
 f_{i,j},\quad & \text{if}\ f_{i,j}\ \text{is in the top-}k_1\ \text{elements of}\ f_i\\
 -\infty,\quad & \text{otherwise}
 \end{aligned}
 \right.
\end{equation}

\noindent \textbf{Relevance Interaction Unit}.
This unit aims to augment user behavior interest by interacting with explicit context from the relevant intra-page items.
As ISU has suppressed the irelevant intra-page items, we deploy cross-attention to utilize the explicit context.
Particularly, we first linearly transform the clicked item and the intra-page items as:
\begin{equation}
 Q_i = xc_i W^Q,\quad K_{i} = p_i W^K,\quad V_{i,j} = p_i W^V
\end{equation}
where $Q,K,V$ represent query, key, and value, respectively, and $W^Q, W^K, W^V \in \mathbb{R}^{D_s \times D_s}$ are transformation matrices.
Then cross-attention is performed to produce the augmented behavior interest:
\begin{equation}
 xc_i^{aug} = (\text{softmax} (\frac{Q_i K_i^{\mathsf{T}}}{\sqrt{D_s}} + f_i) V_i) W^O + xc_i
\end{equation}
where $W^O \in \mathbb{R}^{D_s \times D_s}$ is used to refine the output.
Note that residual learning~\cite{residual} is applied.
In summary, CIU focuses on utilizing intra-page items that might cause user to compare to augment user behavior interest.

\subsection{Implicit Decision-Making Context Modeling}
On the one hand, personalization in pre-ranking makes some homogeneous items included in the candidates.
Before the user make click decisions, these relevant items' co-occurrence implicitly tell us what the user's interests might be.
On the other hand, user's interests are diverse, resulting in a diversity of items in the candidates.
For current target item, it is formidable to leverage this implicit context in such a noisy environment.
Based on these consideration, we want a module to filter out the candidates that are irrelevant to the target and leverage implicit context of the relevant ones.
Fortunately, the proposed CIU meets these requirements, so we repurpose it to refine the representation of the target so as to emphasize its attributes that may activate user interests.
\par Given the candidates $xs=\{xs_1, xs_2, \dots, xs_C\}$ and the target $xt$, ISU first formalizes their relevance $s = \{s_1, s_2, \dots, s_C\}$ as:
\begin{equation}
s_j = \text{MLP} ([xt, xs_j, xt-xs_j, xt \odot xs_j])
\end{equation}
\par The top-$k_2$ relevant candidates are preserved, while the others are suppressed:
\begin{equation}
 s_{j} = \left\{
 \begin{aligned}
 s_{j},\quad & \text{if}\ s_j\ \text{is in the top-}k_2\ \text{elements of}\ s\\
 -\infty,\quad & \text{otherwise}
 \end{aligned}
 \right.
\end{equation}
\par Then RIU deploys cross-attention to aggregate commonalities of the target-relevant candidates and produces a refined target that better activates user interests:
\begin{equation}
\begin{aligned}
 Q = &\ xt W^Q,\quad K = xs W^K,\quad V = xs W^V \\
 xt^{re} & = (\text{softmax} (\frac{Q K^{\mathsf{T}}}{\sqrt{D_t}} + s) V) W^O + xt
\end{aligned}
\end{equation}
where $W^Q, W^K, W^V, W^O$ are now of dimension $D_t \times D_t$.

\subsection{Adaptive Interest Aggregation Unit}
The explicit and implicit contexts are utilized by CIUs to produce the augmented behavior interests and the refined target.
However, the relationship between the two representations has not been modeled, posing two fatal limitations.
First, all the target share the same behavior interests.
Second, all the behavior interests contribute equally when predicting a particular target, but in fact the target only activate partial user interests.
For example, user's earlier clicking on fast foods implies that he may click on the currently displayed burger, while his clicking on flowers provides no useful information.
Therefore, a module that can aggregates user interests according to the target is desired.
\par Note that the distributions of behavior interests and target are different as they are modeled separately with different input features.
Simply applying the proposed CIU can't reduce the difference and leads to inferior interests aggregation.
To relieve this problem, we propose an Adaptive Interest Aggregation Unit (AIAU) shown in~\cref{fig:dcin}(c). Given the augmented behavior interests $xc^{aug} = \{xc^{aug}_1, xc^{aug}_2,\dots,xc^{aug}_S\}$ and the refined target $xt^{re}$, a MLP with two hidden layers is deployed to adaptively align and activate the $i$-th behavior interest according to the target:
\begin{equation}
 xc^{a}_i = \text{MLP} ([xt^{re}, xc^{aug}_i])
\end{equation}
where $xc^{a}_i$ is of dimension $D_a$. Then the aligned interests $xc^a = \{xc^a_1, xc^a_2, \dots,xc^a_S\}$ go through a self-attention layer to capture mutual influence:
\begin{equation}
 xc^m = \text{softmax} (\frac{Q K^{\mathsf{T}}}{\sqrt{D_a}}) V
\end{equation}
where $Q,K,V$ are linearly transformed from $xc^a$:
\begin{equation}
 Q = xc^a W^Q,\quad K = xc^a W^K,\quad V = xc^a W^V
\end{equation}
where $W^Q, W^K, W^V \in \mathbb{R}^{D_a \times D_a}$. An average pooling layer and another two-layer MLP is used to perform final interest aggregation, and the user interest that represents the user's tendency toward the target is produced:
\begin{equation}
 x^s = \text{MLP} (\text{Avg\,Pool} (xc^m))
\end{equation}

\subsection{Optimization Objective}
The aggregated behavior interest is concatenated with the user embedding, the item embedding and the context embedding. The resulting vector is fed into the final MLP to predict CTR:
\begin{equation}
\hat{y} = \text{sigmoid} (\text{MLP} ([x^s, x^u, x^i, x^c]))
\end{equation}
The model is optimized via the negative log-likelihood function:
\begin{equation}
 \ell = -\frac{1}{N} \mathop{\sum}\limits_{k=1}^N (y_k \,log\,\hat{y}_k + (1-y_k)\,log\, (1-\hat{y}_k))
\label{logloss}
\end{equation}
where $N$ denotes the size of the training set, $y \in \{0,1\}$ denotes the label and $\hat{y}$ is the predicted CTR.

\section{Experiments}
%直接说因为没有大规模数据集同时有page行为信息和pre-rank candidate，所以我们构造数据集，并构造了真实数据集来验证我们方法的有效性：1、美团；2、这两个数据集的统计信息如表1所示，具体的构造方式如下：
Since few large scale datasets contain both behavior page information and pre-ranking candidates, we construct a dataset based on the publicly available Avito\footnote{https://www.kaggle.com/c/avito-context-ad-clicks/data.} dataset.
Meanwhile, we collect an industrial dataset, Meituan Waimai Display Ads (MeituanAds for short), from the online service logs of Meituan Waimai, the largest food delivery platform in China.
The statistics of the two datasets are summarized in~\cref{tab:Dataset}, and we detail the two datasets as follows:
\begin{itemize}
 \item Avito. The Avito dataset comes from a random sample of ad logs from avito.ru.
 It contains user search information, such as user\_id, search\_id, and search\_date.
 Each search\_id corresponds to a search page with multiple ads.
 For each user, We rank his search pages in increasing order based on the search\_date, and use the first $T-1$ search pages as the behavior pages, and the ads in the $T$-th search page as the target ads to be predicted.
We construct the candidate set by the co-occurrence rule:
for each target ad, we count the other ads that co-occur with it on the same search page to form a co-occurrence list, and then randomly sample ads from the list to form its candidates.
To avoid data leakage in the training process, we first partition the training set according to users, using only some of them to construct candidates while the others undergo training.
We use 20150428 to 20150518 as the training set, 20150519 as the validation set and 20150520 as the testing set. 
 \item MeituanAds. Since the public dataset does not simultaneously contain behavior pages and pre-ranking candidates, we collected the real behavior pages exposed to the users and the corresponding candidates from the online service logs of Meituan Waimai App from 20220525 to 20220610 as the training set, and collected the data in 20220611 as the validation set and 20220612 as the testing set.
\end{itemize}

\begin{table}
\centering
 \setlength{\tabcolsep}{1.0mm}{
 \begin{tabular}{ccc}
 \toprule
 Dataset & $\text{Avito}^{\dagger}$ & MeituanAds \\
 \midrule
 \# Users & 0.54 million & 0.2 billion \\
 \# Samples & 0.88 million & 5.3 billion \\
 Avg \# Behavior-Pages & 1.9 & 7.3\\
 \# Candidates & 20 & 60\\
 \midrule
 \bottomrule
 \end{tabular}}
 \caption{Statistics of datasets used in our experiments. $\text{Avito}^{\dagger}$ denotes the constructed Avito dataset.}
 \label{tab:Dataset}
\end{table}

\begin{table*}
 \centering
 \setlength{\tabcolsep}{6.0mm}{
 \begin{tabular}{|c||c|c||c|c||}
 \hline
 \multirow{2}*{\text{Model}} & \multicolumn{2}{|c||}{$\text{Avito}^{\dagger}$} & \multicolumn{2}{|c||}{\text{MeituanAds}} \\
 \cline{2-5}
 & LogLoss & AUC & LogLoss & AUC \\ \hline
 DNN & 0.5587 & 0.7756 & 0.1842 & 0.6891 \\
 \cline{1-5}
 DIN~\cite{din} & 0.5496 & 0.7834 & 0.1837 & 0.6936 \\
 DIEN~\cite{dien} & 0.5490 & 0.7830 & 0.1833 & 0.6949 \\
 DFN~\cite{dfn} & 0.5473 & 0.7841 & 0.1833 & 0.6961 \\
 \cline{1-5}
 DSIN~\cite{dsin} & 0.5475 & 0.7847 & 0.1832 & 0.6963 \\
 CIM~\cite{cim} & 0.5459 & 0.7852 & 0.1839 & 0.6960 \\
 RACP~\cite{racp} & 0.5452 & 0.7863 & 0.1830 & 0.6972 \\
 \hline
 \textbf{DCIN (ours)} & \textbf{0.5445} & \textbf{0.7904} & \textbf{0.1825} & \textbf{0.7014} \\ 
 \hline
 \end{tabular}}
 \caption{Performance of different models on datasets. $\text{Avito}^\dagger$ denotes the constructed Avito dataset.}
 \label{tab:results}
\end{table*}

\subsection{Competitors}
We compare our DCIN with the following classic methods. For fairness, all the methods use the same features.
\begin{itemize}
 \item DNN.
 DNN follows an Embedding\&MLP paradigm, i.e., the high-dimensional sparse features are transformed into low-dimensional dense representations, which will be concatenated together and fed into a MLP to predict CTR.
 Note that DNN is also the base of most CTR prediction models.
 \item DIN \& DIEN.
 DIN~\cite{din} and DIEN~\cite{dien} are two pioneering works that model the users' click behaviors, and are successfully deployed in industry.
 The former simply sums the extracted behavior interests, while the latter uses GRU to model interest evolution.
 \item DFN.
 DFN~\cite{dfn} argues that both positive and negative user behaviors can inform inferences about user interests, and proposes a model that learns click sequence and dislike sequance separately.
 Besides, the learned features are utilized to distill the noisy unclick sequence to make better use of information.
 \item DSIN \& RACP. 
 DSIN~\cite{dsin} and RACP~\cite{racp} introduce session structure and page structure to their model, respectively.
 DSIN first models intra-sesssion user interest, and then uses a bidirectional-LSTM to learn interest evolution.
 RACP first models behaviors in each exposure page, and then uses GRU to learn inter-page interest evolution.
 \item CIM. CIM~\cite{cim} couples an impression model and a transformer together to extract user awareness from the candidates generated by upstream relevance filter.
\end{itemize}

\subsection{Implementation Details}
% In the implementation of DNN model, each element of sequential features is regarded as a separate vector.
% For DIN and DIEN, click sequence and intra-page items (organized as unclick sequence) are modeled separately.
% DFN interacts the click and unclick sequences.
% DSIN divides the sequences by date and then models them.

 The outputs of these feature interaction networks are combined with the remaining features, and fed into final MLPs for CTR prediction.
 The final MLPs in all experiments contain two layers with 256 and 128 hidden units.
 We use AdaGrad~\cite{adagrad} to optimize all the networks.
 The hyper-parameters are set as follows:
for the constructed Avito dataset, the length of click sequence $S=5$, the number of intra-page items $P=5$, and the number of candidate items $C=20$;
for MeituanAds dataset, we set the length of click sequence $S=20$, the number of intra-page items $P=10$, and the number of candidate items $C=60$.
 The values of $k$ in explicit/implicit CIUs are selected experimentally (see ablation study for details).
Our models are trained in a large-scale machine learning platform in Meituan.

\subsection{Evaluation Metrics}
We use Logloss in~\cref{logloss} and Area Under Curve (AUC) as our evaluation metrics.
Logloss measures the distance between the predicted probability of the model and the label, the lower the better.
AUC is the most commonly used evaluation metric for CTR prediction task. It measures the probability of a model rank a randomly chosen positive instance higher than a randomly chosen negative one, and has good offline and online consistency. We calculate AUC as follows:
\begin{equation}
 \text{AUC} = \frac{1}{N^{+} N^{-}} \sum_{x^{+} \in \mathbb{D}^{+}} \sum_{x^{-} \in \mathbb{D}^{-}} ( \mathbb{I} (f (x^{+}) > f (x^{-})))
\end{equation}
where $\mathbb{D}^{+}, \mathbb{D}^{-}$ denote positive and negative instances, and $N^{+}, N^{-}$ denote their quantities. $\mathbb{I}$ is the indicator function and $f (\cdot)$ is the CTR prediction function.

\subsection{Results on Datasets}
The quantitative results on the constructed Avito dataset and MeituanAds dataset are summarized in~\cref{tab:results}.
All experiments were repeated 5 times and the averaged results are reported. 
As can be seen from the table, DIN and DIEN are much more effective than DNN because they model user behavior.
DFN gets better results as both click and unclick behaviors are exploited.
It is worth noting that DSIN and RACP, especially the latter, further improve the CTR prediction because of the information structure they introduce.
By modeling and interacting the valuable decision-making contexts, our DCIN achieves 0.0041/0.0042 absolute AUC gain over RACP on the $\text{Avito}^\dagger$/MeituanAds datasets.
Note that for industrial recommendation and advertising systems, 0.001 absolute AUC gain is very significant.

\begin{center}
\begin{table}[ht]
 \centering
 \setlength{\tabcolsep}{8.0mm}{
 \begin{tabular}{cc}
 \toprule
 Models & AUC \\
 \midrule
 DCIN w/o explicit CIU & 0.6986 \\
 DCIN w/o implicit CIU & 0.6991 \\
 DCIN w/o AIAU & 0.6993 \\
 \midrule
 \textbf{DCIN} & \textbf{0.7014} \\
 \bottomrule
 \end{tabular}}
 \caption{Ablation studies of the components. Each component brings significant improvement in AUC, verifying their effectiveness.}
 \label{tab:ablation_CIU}
\end{table}
\end{center}

\subsection{Ablation Study}
To explore the effectiveness of different modules in DCIN, we conduct ablation studies on MeituanAds dataset. All experiments were repeated 5 times and the averaged AUC is reported.

\noindent \textbf{The impact of CIU.}
In DCIN, CIUs are deployed to aggregate explicit and implicit contexts from the relevant items while suppressing the influence from the irrelevant ones.
To verify their effectiveness, we replace the explicit/implicit CIUs with sum pooling.
As shown in~\cref{tab:ablation_CIU}, AUC decreases by 0.0028/0.0023, suggesting that our CIUs are capable to distinguish and utilize the useful explicit/implicit contexts.

\noindent \textbf{The impact of AIAU.}
After augmenting the user behavior interests and refining the target representation, AIAU is deployed to laern their relation to extract target-specific user interest.
To verify AIAU's effectiveness, we test the performance of DCIN without AIAU: we sum pool all the augmented behavior interests, and the resulted vector is fed to the final MLP together with the refined target.
As shown in~\cref{tab:ablation_CIU}, AUC decreases by 0.0021, suggesting that extracting target-specific interest is crucial and the proposed AIAU meets this requirement.

\begin{figure*}[t]
 \centering
 \includegraphics[width=0.8\linewidth]{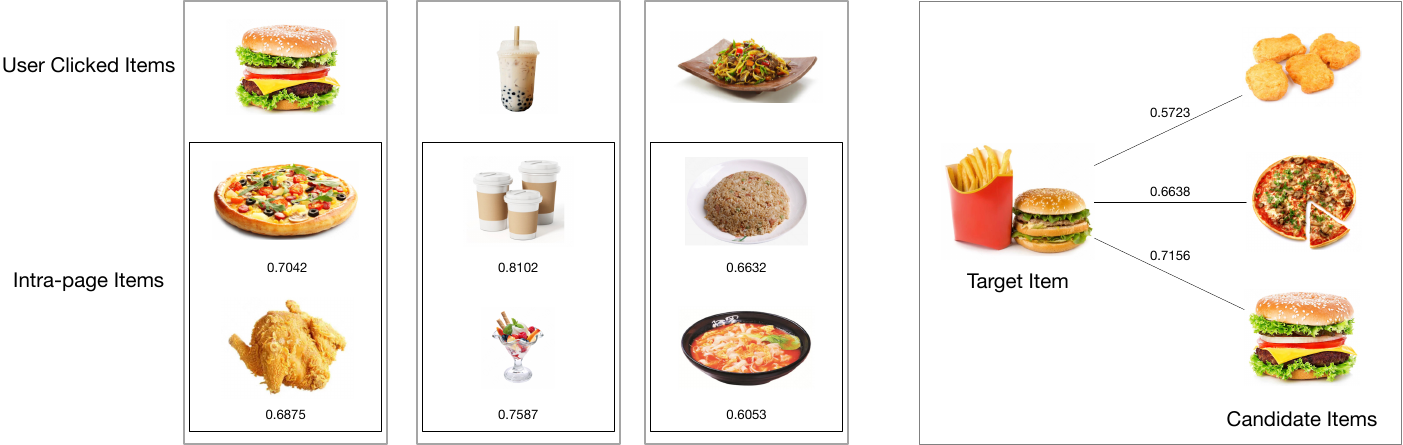}
 \caption{Case study on MeituanAds dataset. We show the relevance score calculated in CIUs in one CTR prediction. The left part corresponds to explicit context modeling, and the right part corresponds to implicit context modeling.}
 \label{fig:case_study}
\end{figure*}

\begin{figure}[ht]
\centering
\includegraphics[width=0.3\textwidth]{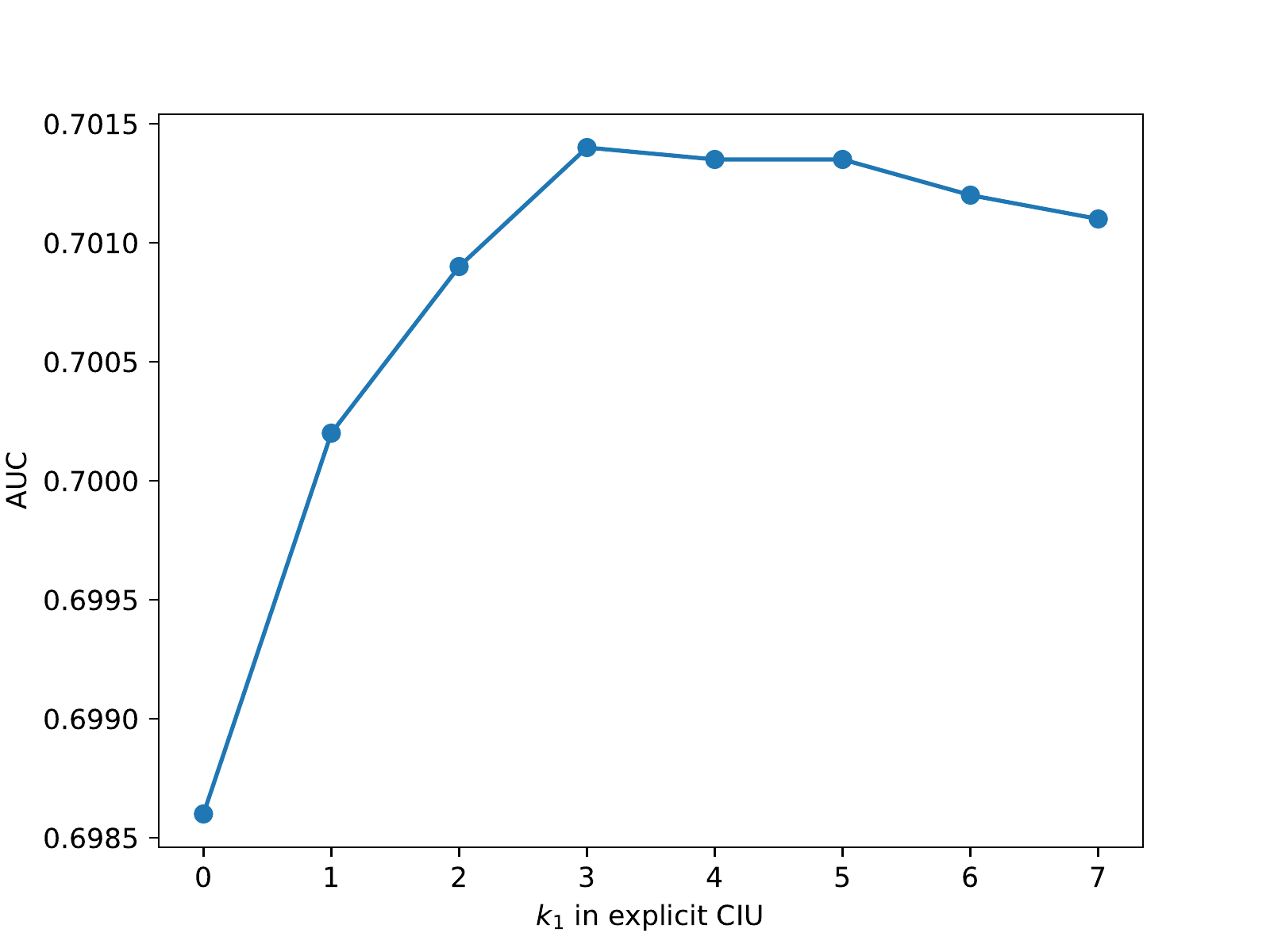}
\caption{Ablation studies of $k_1$ in explicit CIU.}
\label{fig:k_1}
\end{figure}

\begin{figure}[ht]
\centering
\includegraphics[width=0.3\textwidth]{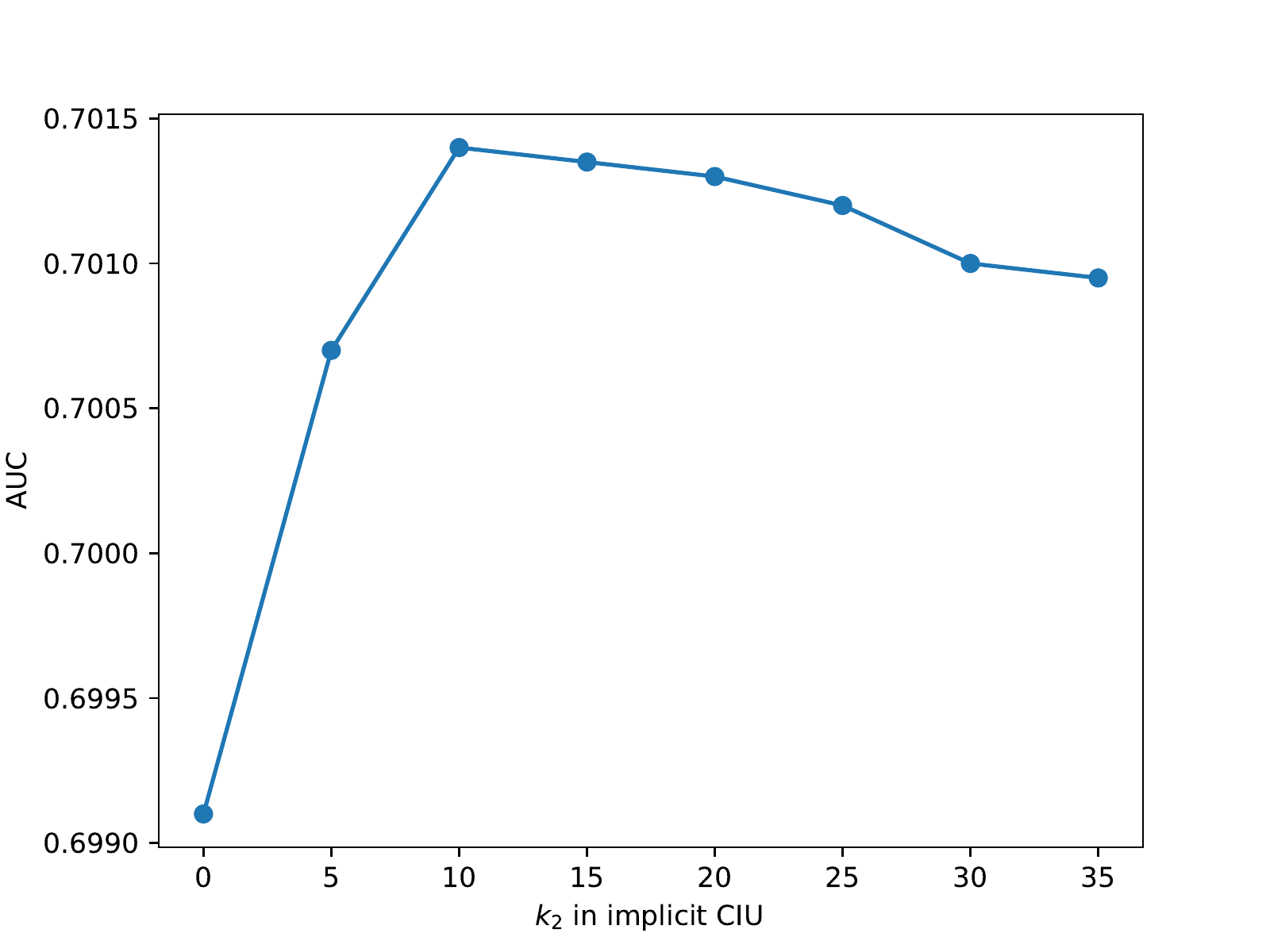}
\caption{Ablation studies of $k_2$ in implicit CIU.}
\label{fig:k_2}
\end{figure}

\noindent \textbf{The impact of hyper-parameter $k$ in CIUs.} CIU preserves only $k$ relevant items while the other irrelevant items are suppressed.
To select the suitable $k$ values, we conduct experiments on explicit/implicit CIUs with different $k$.
As shown in~\cref{fig:k_1} and~\cref{fig:k_2}, when $k_1$ in explicit CIU is set to $3$ and $k_2$ in implicit CIU is set to 10, DCIN performs best.
In~\cref{fig:k_1}, as $k_1$ increases, the model performs better because the informative explicit context is utilized.
However, the performance gradually decreases when increasing $k_1$ further, we attribute this to the introduction of noisy intra-page items.
Similarly, in~\cref{fig:k_2}, setting $k_2=10$ to learn implicit context is optimal on our experiments, and increasing $k_2$ further is negative because the candidates tends to be noisy.

\begin{table}
 \centering
 \setlength{\tabcolsep}{1.0mm}{
 \begin{tabular}{ccccc}
 \toprule
 \centering
 Model & CTR & CPM & GMV\\
 \midrule
 DCIN & +2.9\% & +2.1\% & +1.5\% \\
 \midrule
 \end{tabular}}
 \caption{The performance in real online advertising system.}
 \label{tab:ABtest}
\end{table}

\subsection{Online A/B Testing}
We conduct A/B testing on Meituan Waimai advertising system with 10\% online traffic from 20220705 to 20220711. The following metrics are reported in~\cref{tab:ABtest}: Click-Through Rate (CTR), Cost Per Mille (CPM), and Gross Merchandise Volume (GMV). It is worth noting that our online prediction model has been highly optimized and the improvement in~\cref{tab:ABtest} is significant. Now DCIN has been deployed online in Meituan Waimai ad system, serving the main traffic of hundreds of millions of users.

\subsection{Case Study}
We conduct case study to verify that CIUs are able to select out the most relevant items.
In the left three columns in~\cref{fig:case_study}, the relevance between user clicked items and the corresponding intra-page items are shown.
For the clicked burger, its relevant items are pizza and fried chicken as they are all fast foods.
The right part in~\cref{fig:case_study} shows the relevance between target item and pre-ranking candidate items.
When the target item is the fries burger combo, the fried chicken nuggets, pizza, and another burger in the candidates are activated.
These visualizations show that the proposed CIU is able to identify the most relevant items, thus enabling the effective use of explicit/implicit contexts.         

\section{Conclusion}
In this paper, we emphasize that recent CTR prediction methods do not fully utilize decision-making contexts available in the recommendation and advertising systems and achieve only suboptimal performance.
To alleviate this problem, we introduce the Decision-Making Context Interaction Network (DCIN), which simultaneously model explicit and implicit decision-making contexts in advertising system.
In particular, a Context Interaction Unit is proposed to distinguish and utilize beneficial contexts.
Besides, an Adaptive Interest Aggregation Unit is proposed to aggregate target-specific user behavior interests.
The performance in extensive offline and online experiments demonstrate the effectiveness of our model.

\bibliography{aaai23}

\begin{thebibliography}{24}
\providecommand{\natexlab}[1]{#1}

\bibitem[{Cheng et~al.(2016)Cheng, Koc, Harmsen, Shaked, Chandra, Aradhye,
  Anderson, Corrado, Chai, Ispir et~al.}]{wide_deep}
Cheng, H.-T.; Koc, L.; Harmsen, J.; Shaked, T.; Chandra, T.; Aradhye, H.;
  Anderson, G.; Corrado, G.; Chai, W.; Ispir, M.; et~al. 2016.
\newblock Wide \& deep learning for recommender systems.
\newblock In \emph{Proceedings of the 1st workshop on deep learning for
  recommender systems}, 7--10.

\bibitem[{Chung et~al.(2014)Chung, Gulcehre, Cho, and Bengio}]{gru}
Chung, J.; Gulcehre, C.; Cho, K.; and Bengio, Y. 2014.
\newblock Empirical evaluation of gated recurrent neural networks on sequence
  modeling.
\newblock \emph{arXiv preprint arXiv:1412.3555}.

\bibitem[{Covington, Adams, and Sargin(2016)}]{covington2016deep}
Covington, P.; Adams, J.; and Sargin, E. 2016.
\newblock Deep neural networks for youtube recommendations.
\newblock In \emph{Proceedings of the 10th ACM conference on recommender
  systems}, 191--198.

\bibitem[{Duchi, Hazan, and Singer(2011)}]{adagrad}
Duchi, J.; Hazan, E.; and Singer, Y. 2011.
\newblock Adaptive subgradient methods for online learning and stochastic
  optimization.
\newblock \emph{Journal of machine learning research}, 12(7).

\bibitem[{Fan et~al.(2022)Fan, Ou, Gu, Fu, Li, Bao, Dai, Zeng, Zhuang, and
  Liu}]{racp}
Fan, Z.; Ou, D.; Gu, Y.; Fu, B.; Li, X.; Bao, W.; Dai, X.-Y.; Zeng, X.; Zhuang,
  T.; and Liu, Q. 2022.
\newblock Modeling Users' Contextualized Page-wise Feedback for Click-Through
  Rate Prediction in E-commerce Search.
\newblock In \emph{Proceedings of the Fifteenth ACM International Conference on
  Web Search and Data Mining}, 262--270.

\bibitem[{Feng et~al.(2019)Feng, Lv, Shen, Wang, Sun, Zhu, and Yang}]{dsin}
Feng, Y.; Lv, F.; Shen, W.; Wang, M.; Sun, F.; Zhu, Y.; and Yang, K. 2019.
\newblock Deep session interest network for click-through rate prediction.
\newblock \emph{arXiv preprint arXiv:1905.06482}.

\bibitem[{Guo et~al.(2017)Guo, Tang, Ye, Li, and He}]{deepfm}
Guo, H.; Tang, R.; Ye, Y.; Li, Z.; and He, X. 2017.
\newblock DeepFM: a factorization-machine based neural network for CTR
  prediction.
\newblock \emph{arXiv preprint arXiv:1703.04247}.

\bibitem[{He et~al.(2016)He, Zhang, Ren, and Sun}]{residual}
He, K.; Zhang, X.; Ren, S.; and Sun, J. 2016.
\newblock Deep residual learning for image recognition.
\newblock In \emph{Proceedings of the IEEE conference on computer vision and
  pattern recognition}, 770--778.

\bibitem[{Hochreiter and Schmidhuber(1997)}]{lstm}
Hochreiter, S.; and Schmidhuber, J. 1997.
\newblock Long short-term memory.
\newblock \emph{Neural computation}, 9(8): 1735--1780.

\bibitem[{Lian et~al.(2018)Lian, Zhou, Zhang, Chen, Xie, and Sun}]{xdeepfm}
Lian, J.; Zhou, X.; Zhang, F.; Chen, Z.; Xie, X.; and Sun, G. 2018.
\newblock xdeepfm: Combining explicit and implicit feature interactions for
  recommender systems.
\newblock In \emph{Proceedings of the 24th ACM SIGKDD international conference
  on knowledge discovery \& data mining}, 1754--1763.

\bibitem[{Liao et~al.(2022)Liao, Shi, Wang, Wu, Zhang, Wang, Wang, and
  Wang}]{dpin}
Liao, G.; Shi, X.; Wang, Z.; Wu, X.; Zhang, C.; Wang, Y.; Wang, X.; and Wang,
  D. 2022.
\newblock Deep Page-Level Interest Network in Reinforcement Learning for Ads
  Allocation.
\newblock In \emph{Proceedings of the 45th International ACM SIGIR Conference
  on Research and Development in Information Retrieval}, SIGIR '22,
  2292–2296. New York, NY, USA: Association for Computing Machinery.
\newblock ISBN 9781450387323.

\bibitem[{Liu et~al.(2020)Liu, LU, Zhao, Xu, Peng, Liu, Zhang, Li, Jin, Bao,
  and Yan}]{kalman}
Liu, H.; LU, J.; Zhao, X.; Xu, S.; Peng, H.; Liu, Y.; Zhang, Z.; Li, J.; Jin,
  J.; Bao, Y.; and Yan, W. 2020.
\newblock Kalman Filtering Attention for User Behavior Modeling in CTR
  Prediction.
\newblock In Larochelle, H.; Ranzato, M.; Hadsell, R.; Balcan, M.; and Lin, H.,
  eds., \emph{Advances in Neural Information Processing Systems}, volume~33,
  9228--9238. Curran Associates, Inc.

\bibitem[{Ouyang et~al.(2019)Ouyang, Zhang, Li, Zou, Xing, Liu, and Du}]{dstn}
Ouyang, W.; Zhang, X.; Li, L.; Zou, H.; Xing, X.; Liu, Z.; and Du, Y. 2019.
\newblock Deep spatio-temporal neural networks for click-through rate
  prediction.
\newblock In \emph{Proceedings of the 25th ACM SIGKDD International Conference
  on Knowledge Discovery \& Data Mining}, 2078--2086.

\bibitem[{Pi et~al.(2020)Pi, Zhou, Zhang, Wang, Ren, Fan, Zhu, and Gai}]{sim}
Pi, Q.; Zhou, G.; Zhang, Y.; Wang, Z.; Ren, L.; Fan, Y.; Zhu, X.; and Gai, K.
  2020.
\newblock Search-based user interest modeling with lifelong sequential behavior
  data for click-through rate prediction.
\newblock In \emph{Proceedings of the 29th ACM International Conference on
  Information \& Knowledge Management}, 2685--2692.

\bibitem[{Song et~al.(2019)Song, Shi, Xiao, Duan, Xu, Zhang, and
  Tang}]{autoint}
Song, W.; Shi, C.; Xiao, Z.; Duan, Z.; Xu, Y.; Zhang, M.; and Tang, J. 2019.
\newblock Autoint: Automatic feature interaction learning via self-attentive
  neural networks.
\newblock In \emph{Proceedings of the 28th ACM International Conference on
  Information and Knowledge Management}, 1161--1170.

\bibitem[{Vaswani et~al.(2017)Vaswani, Shazeer, Parmar, Uszkoreit, Jones,
  Gomez, Kaiser, and Polosukhin}]{attn_need}
Vaswani, A.; Shazeer, N.; Parmar, N.; Uszkoreit, J.; Jones, L.; Gomez, A.~N.;
  Kaiser, {\L}.; and Polosukhin, I. 2017.
\newblock Attention is all you need.
\newblock \emph{Advances in neural information processing systems}, 30.

\bibitem[{Wang et~al.(2017)Wang, Fu, Fu, and Wang}]{dcn}
Wang, R.; Fu, B.; Fu, G.; and Wang, M. 2017.
\newblock Deep \& cross network for ad click predictions.
\newblock In \emph{Proceedings of the ADKDD'17}, 1--7.

\bibitem[{Wang et~al.(2020)Wang, Zhao, Jiang, Zhou, Zhu, and
  Gai}]{wang2020cold}
Wang, Z.; Zhao, L.; Jiang, B.; Zhou, G.; Zhu, X.; and Gai, K. 2020.
\newblock Cold: Towards the next generation of pre-ranking system.
\newblock \emph{arXiv preprint arXiv:2007.16122}.

\bibitem[{Xie et~al.(2021)Xie, Ling, Wang, Wang, Xia, and Lin}]{dfn}
Xie, R.; Ling, C.; Wang, Y.; Wang, R.; Xia, F.; and Lin, L. 2021.
\newblock Deep feedback network for recommendation.
\newblock In \emph{Proceedings of the Twenty-Ninth International Conference on
  International Joint Conferences on Artificial Intelligence}, 2519--2525.

\bibitem[{Zheng et~al.(2022)Zheng, Wang, Li, Chen, Liu, Lu, Zhao, Peng, Lin,
  and Shao}]{cim}
Zheng, K.; Wang, L.; Li, Y.; Chen, X.; Liu, H.; Lu, J.; Zhao, X.; Peng, C.;
  Lin, Z.; and Shao, J. 2022.
\newblock Implicit User Awareness Modeling via Candidate Items for CTR
  Prediction in Search Ads.
\newblock In \emph{Proceedings of the ACM Web Conference 2022}, 246--255.

\bibitem[{Zhou et~al.(2020)Zhou, Bian, Wu, Ren, Pi, Zhang, Xiao, Sheng, Mou,
  Luo et~al.}]{can}
Zhou, G.; Bian, W.; Wu, K.; Ren, L.; Pi, Q.; Zhang, Y.; Xiao, C.; Sheng, X.-R.;
  Mou, N.; Luo, X.; et~al. 2020.
\newblock CAN: revisiting feature co-action for click-through rate prediction.
\newblock \emph{arXiv preprint arXiv:2011.05625}.

\bibitem[{Zhou et~al.(2019)Zhou, Mou, Fan, Pi, Bian, Zhou, Zhu, and Gai}]{dien}
Zhou, G.; Mou, N.; Fan, Y.; Pi, Q.; Bian, W.; Zhou, C.; Zhu, X.; and Gai, K.
  2019.
\newblock Deep interest evolution network for click-through rate prediction.
\newblock In \emph{Proceedings of the AAAI conference on artificial
  intelligence}, volume~33, 5941--5948.

\bibitem[{Zhou et~al.(2018)Zhou, Zhu, Song, Fan, Zhu, Ma, Yan, Jin, Li, and
  Gai}]{din}
Zhou, G.; Zhu, X.; Song, C.; Fan, Y.; Zhu, H.; Ma, X.; Yan, Y.; Jin, J.; Li,
  H.; and Gai, K. 2018.
\newblock Deep interest network for click-through rate prediction.
\newblock In \emph{Proceedings of the 24th ACM SIGKDD international conference
  on knowledge discovery \& data mining}, 1059--1068.

\bibitem[{Zhu et~al.(2018)Zhu, Li, Zhang, Li, He, Li, and
  Gai}]{zhu2018learning}
Zhu, H.; Li, X.; Zhang, P.; Li, G.; He, J.; Li, H.; and Gai, K. 2018.
\newblock Learning tree-based deep model for recommender systems.
\newblock In \emph{Proceedings of the 24th ACM SIGKDD International Conference
  on Knowledge Discovery \& Data Mining}, 1079--1088.

\end{thebibliography}

\end{document}